# Optical Continuum Reverberation Mapping of a Candidate IMBH in a Nearby Seyfert 1 Galaxy

Wenwen Zuo[1], Hengxiao Guo[1], Jingbo Sun[1,2], Qi Yuan[1,3], Paulina Lira[4], Minfeng Gu[1], Philip G. Edwards[5], Alok C. Gupta[6], Shubham Kishore[6,7], Jamie Stevens[5], Tao An[1], Zhen-Yi Cai[8,9], Hai-Cheng Feng[10,11,12], Luis C. Ho[13,14], Dragana Ilić[15,16], Andjelka B. Kovačević[15], Sha-Sha Li[10,11,12], Mar Mezcua[17,18], Luka Č. Popović[15,19], Mouyuan Sun[20], Tushar Tripathi[6,7], Vivian U[21], Oliver Vince[19], Jianguo Wang[10,11], Junxian Wang[8,9], Shu Wang[22], Xuebing Wu[13,14], and Zhenya Zheng[1]

[1] Shanghai Astronomical Observatory, Chinese Academy of Sciences, 80 Nandan Road, Shanghai 200030, People's Republic of China; hengxiaoguo@gmail.com, wenwenzuo@shao.ac.cn
[2] University of Chinese Academy of Sciences, 19A Yuquan Road, 100049, Beijing, People's Republic of China
[3] Changchun Observatory, National Astronomical Observatories, Chinese Academy of Sciences, West Hill of Jingyue Lake, Changchun 130117, People's Republic of China
[4] Departamento de Astronomía, Universidad de Chile, Casilla 36D, Santiago, Chile
[5] CSIRO Astronomy and Space Science, PO Box 76, Epping, NSW, 1710, Australia
[6] Aryabhatta Research Institute of Observational Sciences (ARIES), Manora Peak, Nainital 263001, India
[7] Department of Physics, DDU Gorakhpur University, Gorakhpur 273009, India
[8] CAS Key Laboratory for Research in Galaxies and Cosmology, Department of Astronomy, University of Science and Technology of China, Hefei, Anhui 230026, People's Republic of China
[9] School of Astronomy and Space Science, University of Science and Technology of China, Hefei 230026, People's Republic of China
[10] Yunnan Observatories, Chinese Academy of Sciences, 396 Yangfangwang, Guandu District, Kunming 650216, Yunnan, People's Republic of China
[11] Key Laboratory for the Structure and Evolution of Celestial Objects, Chinese Academy of Sciences, Kunming 650216, Yunnan, People's Republic of China
[12] Center for Astronomical Mega-Science, Chinese Academy of Sciences, 20A Datun Road, Chaoyang District, Beijing 100012, People's Republic of China
[13] Department of Astronomy, School of Physics, Peking University, Beijing 100871, People's Republic of China
[14] Kavli Institute for Astronomy and Astrophysics, Peking University, Beijing, 100871, People's Republic of China
[15] University of Belgrade, Faculty of Mathematics, Department of Astronomy, Studentski trg 16, Belgrade, Serbia
[16] Hamburger Sternwarte, Universität Hamburg, Gojenbergsweg 112, D-21029 Hamburg, Germany
[17] Institute of Space Sciences (ICE, CSIC), Campus UAB, Carrer de Magrans, 08193 Barcelona, Spain
[18] Institut d'Estudis Espacials de Catalunya (IEEC), Edifici RDIT, Campus UPC, 08860 Castelldefels (Barcelona), Spain
[19] Astronomical Observatory, Volgina 7, 11060 Belgrade, Serbia
[20] Department of Astronomy, Xiamen University, Xiamen, Fujian 361005, People's Republic of China
[21] Department of Physics and Astronomy, University of California, Irvine, 4129 Frederick Reines Hall, Irvine, CA 92697-4575, USA
[22] Department of Physics & Astronomy, Seoul National University, Seoul 08826, Republic of Korea

## Abstract

To investigate the short-term variability and determine the size of the optical continuum emitting region of intermediate-mass black holes (IMBHs), we carried out high-cadence, multiband photometric monitoring of a Seyfert 1 galaxy J0249−0815 across two nights, together with a one-night single-band preliminary test. The presence of the broad H$\alpha$ component in our target was confirmed by recent Palomar/P200 spectroscopic observations, 23 yr after the Sloan Digital Sky Survey, ruling out the supernovae origin of the broad H$\alpha$ line. The photometric experiment was primarily conducted utilizing four-channel imagers MuSCAT 3 and 4 mounted on 2 m telescopes within the Las Cumbres Observatory Global Telescope Network. Despite the expectation of variability, we observed no significant variation (<1.4%) on timescales of 6–10 hr. This nondetection is likely due to substantial host galaxy light diluting the subtle active galactic nucleus (AGN) variability. Additionally, we cannot rule out that the target was in a relatively quiescent state without intranight variability during our monitoring, owing to the stochastic nature of AGN variations. To enhance the possibility of detecting subtle variability signals and lag in future IMBH reverberation campaigns, it may be beneficial to select targets with a higher AGN-to-host flux ratio, and conduct dual-band preliminary tests and tailored simulations.

*Unified Astronomy Thesaurus concepts:* Intermediate-mass black holes (816); Black hole physics (159); Reverberation mapping (2019)

## 1. Introduction

Accurate measurements of the mass of intermediate-mass black holes (IMBHs; $M_{\rm BH} \sim 10^{2-6} M_\odot$) and a comprehensive census are crucial in unraveling essential clues regarding the formation of the seeds of black holes (BHs) and supermassive black holes (SMBHs) during the early stages of the Universe (see the reviews by M. Mezcua 2017; J. E. Greene et al. 2020; K. Inayoshi et al. 2020; M. Volonteri et al. 2021; A. E. Reines 2022). Continuum reverberation mapping (RM) offers a unique way to indirectly resolve the exceedingly small scale of the accretion disk, which allows us to directly estimate the BH mass via the scaling relation of the continuum emitting size and the broad emission-line size (S. Wang et al. 2023), validate the standard disk model (N. I. Shakura & R. A. Sunyaev 1973), and test the variability mechanism of active galactic nuclei (AGNs; E. M. Cackett et al. 2021).

Like in luminous AGNs, the optical variability is an efficient way to detect IMBH AGNs. V. F. Baldassare et al. (2018) undertook a systematic exploration of the Sloan Digital Sky Survey (SDSS) Stripe 82 light curves encompassing approximately 28,000 galaxies, and identified 135 low-mass galaxies exhibiting AGN-like variability. Taking advantage of the High Cadence Transient Survey (F. Förster et al. 2016), J. Martínez-Palomera et al. (2020) proposed the Search for Intermediate-mass Black Holes in Nearby Galaxies campaign, aiming to detect IMBHs utilizing the optical variability method. They preliminarily confirmed 22 AGNs in nearby galaxies based on the Baldwin, Phillips, and Terlevich diagrams (J. A. Baldwin et al. 1981). The space-based Transiting Exoplanet Survey Satellite (TESS) has also been helpful in identifying AGNs through optical variability, leveraging its exceptional cadence and precision (C. J. Burke et al. 2020; H. P. Treiber et al. 2023). Notably, NGC 4449 and NGC 4395 have been identified as two high-confidence IMBH candidates through the selection from TESS. Finally, two-band, intranight photometric monitoring with relatively low cadence has also been designed to search the weak variability in IMBH candidates (L. Shin et al. 2022). Although all these efforts increase the number of IMBH candidates, the principal challenge in conducting IMBH-RM lies in the extremely weak variability exhibited by low-mass BHs, which tends to be overshadowed by their host galaxies.

NGC 4395, an IMBH prototype, is a nearby ($z = 0.001$, $D_L = 4.3$ Mpc), bulgeless spiral dwarf galaxy, which contains the least luminous known Seyfert 1 nucleus exhibiting broad emission lines and rapid optical/X-ray variability (A. V. Filippenko & W. L. W. Sargent 1989; L. C. Ho et al. 1993; P. Lira et al. 1999). Over the past two decades, attempts have been made to measure broad-line and continuum lags (e.g., A. V. Filippenko & L. C. Ho 2003; L.-B. Desroches et al. 2006; D. T. Cameron et al. 2012; H. Edri et al. 2012; I. M. McHardy et al. 2016; H. Cho et al. 2021) although the measured lags are usually too coarse to draw any robust conclusions. However, more recently, photometric broad-line and continuum RM have been successfully carried out (J.-H. Woo et al. 2019; H. Cho et al. 2020, 2021; J. W. Montano et al. 2022; I. M. McHardy et al. 2023). These yielded a precise estimate of the BH mass at $\sim 1.7 \times 10^4\ M_\odot$ and constrained the accretion disk size, making NGC 4395 the first reverberation-confirmed IMBH.[23] These successful campaigns highlight the significance of employing high-cadence photometric RM to capture extremely short continuum and broad-line time lags in low-mass BHs, thus paving the way for applying RM to other IMBHs.

Based on previous experience with NGC 4395, we propose to carry out continuum RM on other IMBH candidates. Our initial sample selection comprises a few dozen of the most robust broad-line IMBH candidates from previous studies (e.g., J. E. Greene & L. C. Ho 2007; X.-B. Dong et al. 2012; H.-Y. Liu et al. 2018). We will refine this sample based on spectral characteristics, both long-term and short-term variability, and multiwavelength data (e.g., radio and X-ray), aiming to concentrate on the 10−20 best candidates for future RM campaigns using 2−10 m telescopes. Our strategy primarily involves high-precision, high-cadence photometric monitoring to detect the subtle variabilities and short lags of IMBHs over several nights per target. The aim is to detect both the continuum and broad-line lags with photometric monitoring, enabling constraints on the accretion disk size and accurate measurement of the BH mass. Continuum RMs of a sample of IMBH candidates will provide a unique opportunity to test the standard thin disk model and the lamppost X-ray reprocessing model (J. H. Krolik et al. 1991; E. M. Cackett et al. 2007) in low-mass BHs on timescales of days, taking advantage of the short lags of IMBHs. In addition, the derived BH mass will facilitate an examination of the applicability of empirical relationships in the low-mass regime, such as the $M-\sigma$ relation (J. Kormendy & L. C. Ho 2013), $R-L$ relation (M. C. Bentz et al. 2013), $M-\tau$ relation (C. J. Burke et al. 2021; Z.-B. Su et al. 2024), and the fundamental plane of BH activity (A. Merloni et al. 2003).

In this work, we report our time series results from the Las Cumbres Observatory Global Telescope (LCOGT; T. M. Brown et al. 2013) and Devasthal Fast Optical Telescope (DFOT)[24] light curves for SDSS J024912.86−081525.6 (hereafter J0249 −0815), which serves as the first experimental test beyond NGC 4395. We describe the technical details of source selection, observations, and data analysis in Section 2. The continuum lags are estimated in Section 3. We present the results and a discussion in Section 4. Finally, we draw our conclusions in Section 5. Throughout this paper, we use the ΛCDM cosmology, with $H_0 = 70.0$ km s$^{-1}$ Mpc$^{-1}$ and $\Omega_m = 0.3$.

## 2. Target Selection, Observations, and Data Reduction

### 2.1. Target Selection

J0249−0815 is a candidate IMBH residing within a dwarf galaxy at $z = 0.0297$ or $D_L \sim 130$ Mpc shown in Figure 1. This IMBH candidate was initially identified by J. E. Greene & L. C. Ho (2004) from the First Data Release of SDSS. The stellar mass is $9.0 \times 10^9\ M_\odot$ obtained from the NASA-Sloan Atlas.[25] The dust-corrected color index of $u - r$ is 1.9, making it a green valley galaxy (K. Schawinski et al. 2014). High-resolution spectroscopic observations were conducted in 2004 using the Keck Echellette Spectrograph and Imager (A. J. Barth et al. 2005; T. Xiao et al. 2011). The measured full width at half-maximum (FWHM) of the broad Hα component is 702 km s$^{-1}$ (corrected for instrumental broadening), which yields a single-epoch BH mass of $2.09 \times 10^5\ M_\odot$ and Eddington ratio $\lambda_{\rm Edd} = L_{\rm bol}/L_{\rm Edd} = 0.31$ based on the broad Hα line, consistent with the results obtained from SDSS spectrum. The stellar velocity dispersion is $53 \pm 3$ km s$^{-1}$, supporting the IMBH hypothesis.

We selected this source as an initial target to carry out photometric continuum RM based on several considerations: (1) the brightness of the galaxy is well suited for monitoring with a 2 m telescope; (2) the anticipated continuum lags are sufficiently small (1.6–3.3 hr, see Section 3 and Figure 3) for effective intranight monitoring based on the relations of lag with respect to the BH mass and the AGN continuum luminosity at 5100 Å; (3) the presence of a prominent broad Hα component, combined with an appropriate redshift, makes it ideal for subsequent broad-line RM using broadband +narrowband filters to measure BH mass.

---

[23] The BH mass of UGC 06728 is $(7.1 \pm 4.0) \times 10^5\ M_\odot$ according to previous spectroscopic RM (M. C. Bentz et al. 2016). However, the Hβ lag of $1.4 \pm 0.8$ days is not well confined due to the relatively low cadence of the light curves. Therefore, we did not consider it as a reliable RM-confirmed IMBH.

[24] https://www.aries.res.in/facilities/astronomical-telescopes/130cm-telescope
[25] http://www.nsatlas.org

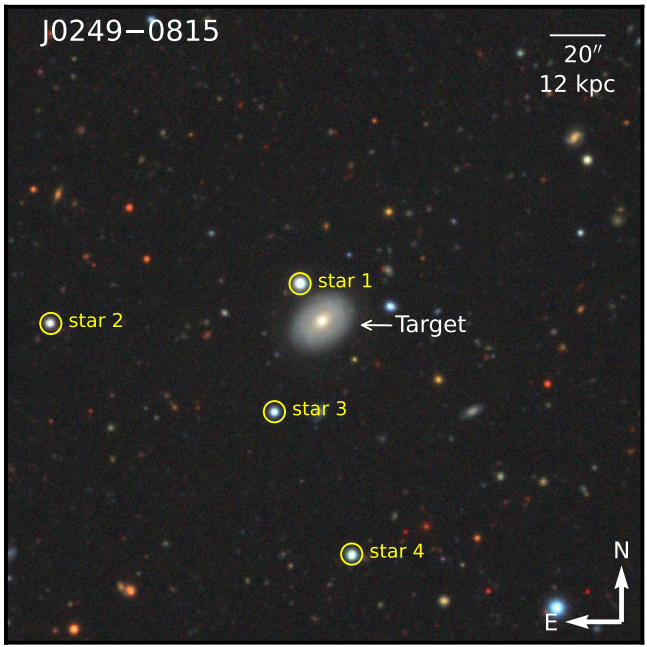

**Figure 1.** A *gri* combined image of J0249−0815 at $z = 0.0297$ from DECaLS with an FOV of $4' \times 4'$. The positions of the target and four reference stars are denoted.

### 2.2. Spectroscopic Observations

To validate the broad Hα component of J0249−0815 over a timescale of 20 yr, we acquired an additional spectrum using the Double Spectrograph (DBSP) mounted on the Hale 5.1 m telescope (P200) at the Palomar Observatory on 2024 January 11. The dichroic D55 split the light beam into red and blue channels at 5500 Å, with a slit width of $1\farcs5$ chosen based on the average seeing conditions at Palomar Observatory. We opted for the 600 lines mm$^{-1}$ (3780 Å blaze) grating on the blue camera and the 316 lines mm$^{-1}$ (7150 Å blaze) grating on the red camera. This configuration provides wavelength coverage spanning ∼4000 to 10000 Å, with a spectral resolution ranging from ∼700 to 1300.[26] The total exposure time of 1800 s was split into two 900 s to eliminate potential cosmic rays. Data processing follows the standard IRAF routine using PyPeIt (J. Prochaska et al. 2020). The processing steps encompass bias subtraction, flat correction, cosmic-ray removal, wavelength/flux calibration, and telluric line correction.

### 2.3. Spectral Fitting

Employing the quasar fitting code PyQSOFit (H. Guo et al. 2018; Y. Shen et al. 2019), we performed the host decomposition for the P200/DBSP and SDSS spectra using the principal component analysis (PCA) based on the galaxy/quasar Eigenspectra obtained from C. W. Yip et al. (2004a, 2004b), as shown in Figure 2. After subtracting the host component, we specifically performed the local fit for the Hα region from 6500 to 6750 Å. A single power law is used to model the continuum emission. Note that we masked the emission-line region when subtracting the host component and assumed all the emission lines in the Hα+[N II] complex are from the AGN, as it is difficult to truly disentangle them.

---

[26] https://sites.astro.caltech.edu/palomar/observer/200inchResources/dbspoverview.html#angle

Although the narrow emission may demonstrate variability on timescale over 20 yr, especially for low-luminosity and low-mass AGNs (B. M. Peterson et al. 2013), we carefully examined the narrow lines and did not find any significant intrinsic variation given different observing conditions and instrumental configurations (see more details in the Appendix). As the continuum and [O III] are already well matched in P200 and SDSS spectra, we have not rescaled the relative flux of two epochs.

Similar to L. C. Ho et al. (1997) and J. E. Greene & L. C. Ho (2004), to break the fitting degeneracy of the Hα+[N II] complex, we first obtained the best-fit profiles of the [S II] doublet and then scaled this narrow component shape ($< 600$ km s$^{-1}$) to that of Hα and [N II]. Specifically, we utilized two Gaussians to model the [S II] doublet, tying their line widths and central wavelengths while allowing the flux ratios to vary. Based on our tests, extra Gaussians to account for outflows in the [S II]-emitting region (J. Kovačević-Dojčinović et al. 2022) were not necessary for our target. Next, we scaled the obtained single Gaussian to fit narrow components of Hα and [N II] since their emitting environments (cloud density and ionization parameter) are similar. The flux ratio of the [N II] doublet was fixed to the theoretical value of 2.96 (D. E. Osterbrock & G. J. Ferland 2006). To accurately model the broad Hα component, we began our analysis with a single broad Gaussian component ($\geqslant 600$ km s$^{-1}$), introducing additional components only if they resulted in a substantial decrease in $\chi^2$ of at least 20%. Ultimately, two Gaussian fits were used for the broad Hα component of the SDSS spectrum, while a single Gaussian fit proved sufficient for the P200 spectrum.

### 2.4. Photometric Monitoring

On 2023 November 9 and 13, we conducted observations of J0249−0815 using the MuSCAT 3 and 4 cameras (N. Narita et al. 2020), mounted on the 2 m Faulkes North and South Telescopes (FTN and FTS). These telescopes are respectively situated at Haleakalā Observatory in Hawaii and Siding Spring Observatory in Australia, both of which are part of the LCOGT network.

The MuSCAT camera functions as a four-channel simultaneous imager, simultaneously observing in the $g'$, $r'$, $i'$, $z_s$ bands (hereafter referred to as *griz*). Its field of view (FOV) spans $9\farcm1 \times 9\farcm1$ with a pixel resolution of $0\farcs267$. To precisely measure the anticipated short interband lags, possibly just a few minutes, we employed the fast readout mode with a readout time of 6 s. For the *giz* bands, exposure times were fixed at 100 s. To mitigate the risk of observation interruption due to loss of guidance with the facility guiding system, we implemented the auto-guiding mode (i.e., the "guide-off" mode), which was executed using 25 s exposures on the *r*-band camera, as NGC 4395 in J. W. Montano et al. (2022).

During the observation on 2023 November 9 (Night 1), we executed continuous monitoring across all four bands, spanning a total duration of ∼8 hr, using the MuSCAT 3. The seeing ranges from $1\farcs3$ to $2\farcs5$, with a median value of $1\farcs7$. Subsequently, another run was carried out on 2023 November 13 (Night 2). Unfortunately, monitoring on MuSCAT 3 was interrupted after ∼3 hr due to the cloudy conditions in Hawaii. After a 3 hr gap, the monitoring was recovered but shifted to MuSCAT 4 for the remaining 4.5 hr. The median seeing was $2\farcs4$ with a standard deviation of $0\farcs27$ for the Night 2.

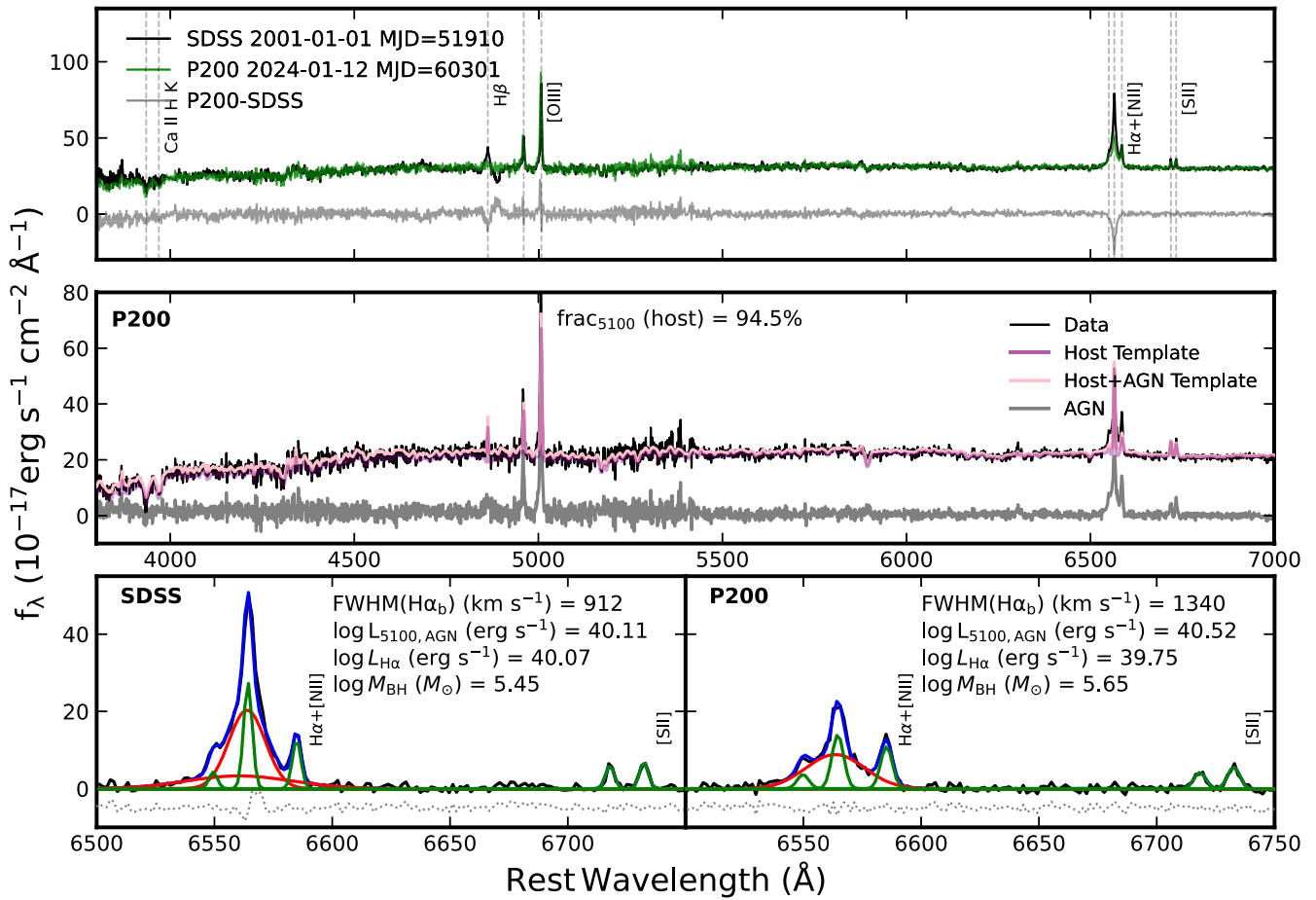

**Figure 2.** Upper panel: spectra of J0249−0815 from SDSS and Palomar P200 in the rest frame. The absolute flux of two original spectra is shifted upward by $1 \times 10^{-16}$ erg s$^{-1}$ cm$^{-2}$ Å$^{-1}$ for better demonstration. Middle panel: host-AGN decomposition with a PCA method. Lower panel: spectral fitting for the continuum-subtracted H$\alpha$ emission-line region. The blue line is the overall model consisting of Gaussian profiles: red lines are broad components for the H$\alpha$ line, and green lines represent narrow emission lines, respectively. All these fits are based on PyQSOFit.

In addition to the two-night LCOGT multiband monitoring, we conducted a preliminary V-band test for this target on 2023 October 13 (Night 0), using the 1.3 m DFOT telescope in India (R. Sagar et al. 2011). Its FOV is 18′ × 18′ with a pixel scale of 0″.535. The single-epoch exposure is set to 50 s throughout the 6.8 hr monitoring on a clear night, with a median seeing of 1″.1.

### 2.5. Image Differencing

Image data collected over two nights from the LCOGT network were processed using the BANZAI reduction pipeline.[27] This comprehensive pipeline includes several critical steps: identifying and masking bad pixels, subtracting bias and dark frames, applying flat-field corrections, and performing astrometric calibration.

After preprocessing, we extracted the light curves for our target and comparison stars through image differencing analysis following R. Kessler et al. (2015). The steps of our image differencing procedure are outlined below:

1. Conducting an initial review of all images to identify and exclude any problematic frames, such as those affected by loss of guiding or significant disruptions caused by strong winds. This evaluation resulted in the removal of 34 frames, approximately 1% of the total data set in four bands.

2. Identification of bright field stars within the FOV was carried out to measure the point-spread function (PSF) kernel for each frame. This process helped in selecting the sharpest image to serve as a reference for subsequent alignment and differencing tasks.

3. We utilized the selected reference image as a benchmark and aligned all single epochs to the reference, employing the Astroalign Python module (M. Beroiz et al. 2020). Although the BANZAI pipeline provides astrometric calibration, Astroalign serves as an additional verification. It operates independently of World Coordinate System (WCS) information and is capable of correcting subtle displacements that may be overlooked by standard WCS-based methods. It identifies and matches three-point asterisms across images to execute the most precise linear transformations.

4. We differenced each frame relative to the reference after convolving the reference to match the PSF size of each frame using HOTPANTS (A. Becker 2015) based on the algorithm in C. Alard & R. H. Lupton (1998) and C. Alard (2000). It derives an optimal kernel solution from a simple least-squares analysis using all the pixels of both scientific image and reference. Simultaneously, it fits

---
[27] https://lco.global/documentation/data/BANZAIpipeline/

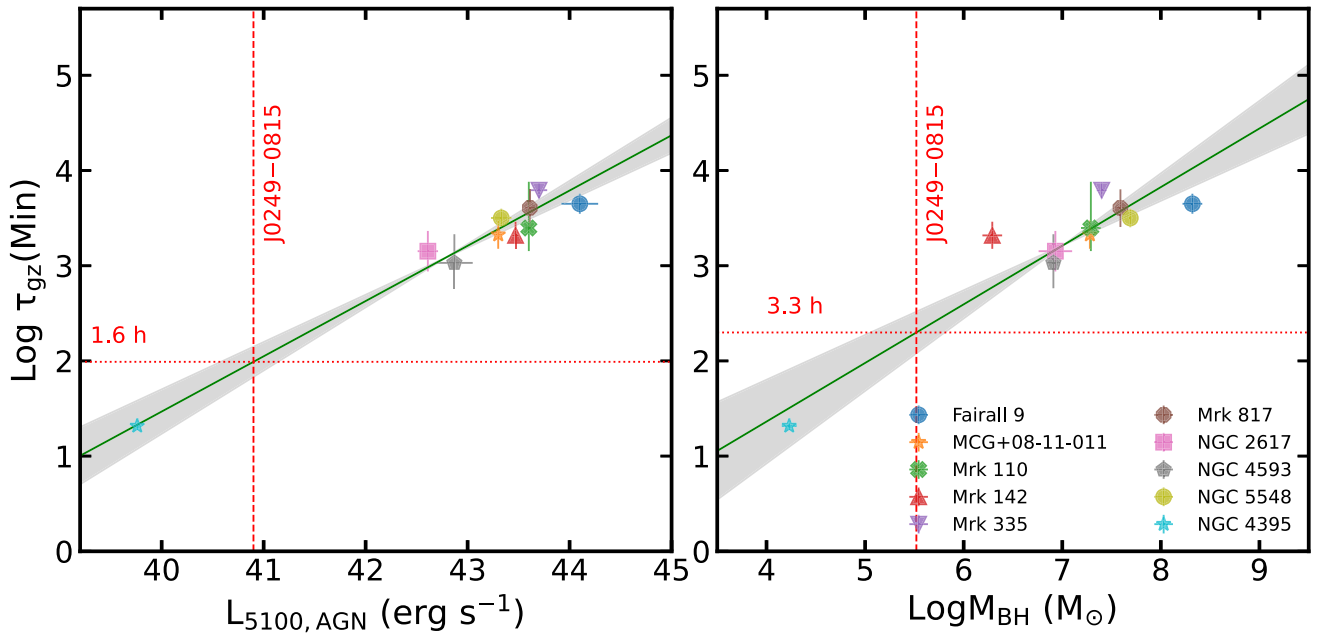

**Figure 3.** Scaling of the $g-z$ continuum lag as a function of AGN luminosity at 5100 Å (left panel) and BH mass (right panel), for 10 nearby AGNs having multiband continuum RM spanning the $g$ to $z$ bands. The best linear fits are shown in green lines with $1\sigma$ uncertainty (light gray region). The vertical red dashed lines correspond to the AGN continuum luminosity at 5100 Å obtained from spectral decomposition and single-epoch BH mass estimates based on the H$\alpha$ component, respectively. The horizontal red dotted lines in both panels indicate the predicted continuum lags ($\sim$1.6 to 3.3 hr) between the $g$ and $z$ bands in J0249−0815.

the differential background variation and removes the background of each frame.

5. To extract the light curves, we performed the force aperture photometry using the Photutils Python package (L. Bradley et al. 2017) for our target and four surrounding comparison stars simultaneously. An aperture size of 15 pixels (4″.005, 1 pixel = 0″.267) was used to accommodate the seeing conditions. Additionally, aperture photometry was performed on the original reference image to determine the initial magnitudes of both the target's nucleus and the comparison stars. Subsequently, absolute flux calibration was carried out using comparison star 1 as a reference.

To ensure the reliability of our pipeline, we replicated the analysis of NGC 4395, as presented in J. W. Montano et al. (2022), using our methodology. Our efforts successfully reproduced the time lags reported in their study, confirming the accuracy of our approach.

### 3. Lag Prediction

Successful continuum RM campaigns, as evidenced in previous studies, such as NGC 5548 (M. M. Fausnaugh et al. 2016) and Mrk 817 (E. Kara et al. 2021) within the Space Telescope and Optical Reverberation Mapping project, have clearly demonstrated the expected positive correlation between lag and wavelength. However, it is noteworthy that the size of the optical emitting region typically exceeds predictions made by the standard thin disk model. To reconcile this disk size problem, various accretion disk models (E. Gardner & C. Done 2017; Z.-Y. Cai et al. 2018; M. Sun et al. 2020; E. S. Kammoun et al. 2021) and additional components, such as diffuse continuum emission (K. T. Korista & M. R. Goad 2019; H. Guo et al. 2022a, 2022b), have been proposed.

In total, approximately 10 continuum lags of SMBHs (E. M. Cackett et al. 2021) have been robustly measured (see Figure 3). At the low-mass regime, the measured minute-level continuum lags in NGC 4395 provide a tight constraint on the relationships at the low-mass and low-luminosity ends. Furthermore, these lag measurements significantly improved our ability to predict continuum lags for other IMBH candidates, thereby guiding the design of tailored continuum RM campaigns.

By collecting all the robust continuum lags reported in the literature and converting them into $g-z$ lags ($\tau_{gz}$), we compiled a sample of 10 local AGNs ($0.001 < z < 0.047$) having high-quality $g$- through $z$-band data from intensive disk RM campaigns: Fairall 9 (J. V. Hernández Santisteban et al. 2020), MCG+08-11-011 (M. M. Fausnaugh et al. 2018), Mrk 110 (F. M. Vincentelli et al. 2021), Mrk 142 (E. M. Cackett et al. 2020), Mrk 335 (E. Kara et al. 2023), Mrk 817 (E. Kara et al. 2021), NGC 2617 (M. M. Fausnaugh et al. 2018), NGC 4593 (E. M. Cackett et al. 2018), NGC 5548 (M. M. Fausnaugh et al. 2016), and NGC 4395 (J. W. Montano et al. 2022). All the values of $\tau_{gz}$ are obtained from J. W. Montano et al. (2022), except for Mrk 335, which is newly added from E. Kara et al. (2023). The continuum luminosities at 5100 Å and BH masses are derived from the M. C. Bentz & S. Katz (2015) catalog.

We then performed the linear fit using the Linmix Python package (B. C. Kelly 2007) for these relations between the $g-z$ lags and $L_{5100,\mathrm{AGN}}$, as well as the BH mass. Two relations are shown as follows:

$$\log(\tau_{gz}/\mathrm{min}) = (0.58 \pm 0.08)\log(L_{5100,\mathrm{AGN}}/\mathrm{erg\ s^{-1}}) - (21.76 \pm 3.72) \quad (1)$$

$$\log(\tau_{gz}/\mathrm{min}) = (0.62 \pm 0.14)\log(M_{\mathrm{BH}}/M_\odot) - (1.10 \pm 0.98). \quad (2)$$

Given the measured AGN continuum luminosity and BH mass of our target, the predicted $\tau_{gz}$ ranges from $\sim$1.6 to 3.3 hr (see Figure 3), detectable within our intranight monitoring.

**Table 1**
The Properties of the IMBH Candidate Derived from Spectral Decomposition

| Name | z | Instrument | MJD | $\log L_{5100,\mathrm{AGN}}$ (erg s$^{-1}$) | $\log L_{\mathrm{H}\alpha}$ (erg s$^{-1}$) | FWHM$_{\mathrm{H}\alpha}$ (km s$^{-1}$) | $\log M_{\mathrm{BH}}$ ($M_\odot$) | $f_{\mathrm{host}}$ |
|---|---|---|---|---|---|---|---|---|
| J0249−0815 | 0.0297 | SDSS | 51910 | 40.11 ± 0.17 | 40.07 ± 0.17 | 912 ± 61 | 5.45 ± 0.08 | 98.3% |
|  |  | P200 | 60301 | 40.52 ± 0.08 | 39.75 ± 0.08 | 1340 ± 65 | 5.65 ± 0.04 | 94.5% |

## 4. Results and Discussion

### 4.1. Ruling Out the Supernovae Origin of the Broad Hα

To evaluate the origin of the broad Hα component, we re-performed the spectroscopic observations of J0249−0815 using the Palomar/P200 telescope. The top panel of Figure 2 illustrates the overall spectral variations between the SDSS and P200 spectra spanning over a 20 yr period. The continuum remains roughly consistent between the two epochs. The primary difference is the total Hα flux caused by broad-line variation.

The middle panel of Figure 2 demonstrates that the host galaxy is the primary contributor to the total light, accounting for ∼95% of the emission at 5100 Å in the P200 spectrum. The presence of Ca II K H absorption lines suggests that the host galaxy comprises a relatively older stellar population. The host-subtracted component exhibits a typical spectrum of an AGN, with a blue power-law-like continuum.

The bottom panel of Figure 2 highlights the Hα complex in 2001 and 2024, revealing the broad Hα component has remained prominent over 20 yr. The SDSS spectral fitting results in an FWHM of 912 ± 61 km s$^{-1}$ and $L_{\mathrm{H}\alpha}$ = 40.07 ± 0.17 erg s$^{-1}$ for a broad Hα component. The P200 yields an FWHM of 1340 ± 65 km s$^{-1}$ and $L_{\mathrm{H}\alpha}$ = 39.75 ± 0.08 erg s$^{-1}$ for a broad Hα component, consistent with the line breathing model (H. Guo et al. 2020; S. Wang et al. 2023). Based on the empirical relation (J. E. Greene & L. C. Ho 2005), the estimated BH masses are both around $10^{5.5} M_\odot$ (see Table 1).

Despite the Hα luminosity has decreased by a factor of ∼2 (0.3 dex), it still strongly supports its origin in AGN activity, rather than being a consequence of shock interactions from supernovae (SNe) with circumstellar material because no SNe-induced broad Hα luminosity to date has been observed to persist above $10^{39}$ erg s$^{-1}$ after a decade of fading (V. F. Baldassare et al. 2016; N. Smith et al. 2017).

### 4.2. The Intranight Variability of J0249−0815

The original light curves of J0249−0815 in the griz bands are depicted in Figure 4. The median value and the standard deviation of the two-night light curves in the griz bands are (2.87 ± 0.04, 3.47 ± 0.06, 3.42 ± 0.03, and 3.24 ± 0.03) × $10^{-16}$ erg s$^{-1}$ cm$^{-2}$ Å$^{-1}$, respectively. Unlike NGC 4395, our target did not show any clear variability trends within two-night monitoring, with the exception of minor fluctuations toward the end of the g band and at the beginning of the z-band light curves, as observed through visual inspections. Given that these fluctuations are not present in other bands, we suggest they might simply be photometric artifacts arising from the image differencing process. The scatter of the original r-band light curve is relatively larger relative to other bands due to the shortest single-epoch exposure time. The typical uncertainties for the griz bands are 0.01, 0.02, 0.01, and 0.01 mag, respectively, given the nucleus magnitudes of 18.05, 17.24, 16.88, and 16.63 mag on dark nights with an aperture size of 4.″005.

Following S. Vaughan et al. (2003), we calculate the fractional variability $F_{\mathrm{var}}$ and the uncertainty of a light curve as follows:

$$F_{\mathrm{var}} = \frac{1}{\langle f(t) \rangle}\sqrt{\Delta^2 - \langle \sigma^2 \rangle} \quad (3)$$

$$\Delta = \frac{1}{N-1}\sum_{i=1}^{N}[f(t_i) - \langle f(t) \rangle]^2 \quad (4)$$

$$\sigma^2_{F_{\mathrm{var}}} = \left(\sqrt{\frac{1}{2N}}\frac{\langle \sigma^2 \rangle}{\langle f(t)^2 \rangle F_{\mathrm{var}}}\right)^2 + \left(\sqrt{\frac{\langle \sigma^2 \rangle}{N}}\frac{1}{\langle f(t) \rangle}\right)^2 \quad (5)$$

where $f(t_i)$ is the flux of the light curve at epoch $i$, $N$ is the number of observations, and $\sigma_i$ is the associated uncertainty.

Table 2 presents the variability properties of the light curves. Columns (2–4) display the fractional variability for Night 1, and the first and second halves of Night 2, respectively. The g-band light curve for the second half of Night 2 and the i-band light curve for Night 1 exhibit very weak variability, with $F_{\mathrm{var}}$ values of 1.4 ± 0.1% and 0.3 ± 0.1%, respectively. However, the other segments show no detectable variability, as their $\Delta^2$ values are less than $\langle \sigma^2 \rangle$ by ranges from 9.0% to 56.7%, which are denoted as $F_{\mathrm{var}} = 0$ in Table 2. Consequently, the LCOGT data quality constrains the nonvariability hypothesis with a precision down to 1.4% in the optical bands for our target.

Based on the data from the 1.3 m DFOT telescope in India, the V-band differential light curve was generated using the same pipeline and an identical aperture size as LCOGT observations. We further corrected the filter difference between the V and g bands based on the P200 spectrum. As shown in Figure 5, the preliminary test light curve from DFOT exhibits no robust variability. This slight offset between the corrected V and g bands is likely introduced by different photometric conditions. However, we cannot completely rule out the possibility that this discrepancy may also reflect some intrinsic AGN variability.

To account for the undetected variability for our IMBH candidate on three different nights, we propose several possibilities:

1. *Host contamination:* To the low-mass end, the AGN contribution diminishes and gradually becomes overshadowed by the host light, challenging the variability detection (e.g., V. F. Baldassare et al. 2018; C. J. Burke et al. 2022; H. P. Treiber et al. 2023). Compared to NGC 4395 at only 4.3 Mpc faraway, our newly selected target resides at a significantly larger distance ($D_L \sim$130 Mpc) and is ∼2 mag fainter in the g band. Moreover, its projected scale on the sky is around 1/36 that of NGC 4395, increasing the difficulty to separate AGN and host components either on image or spectrum. Based on the data from the same facility LCOGT, the photometric

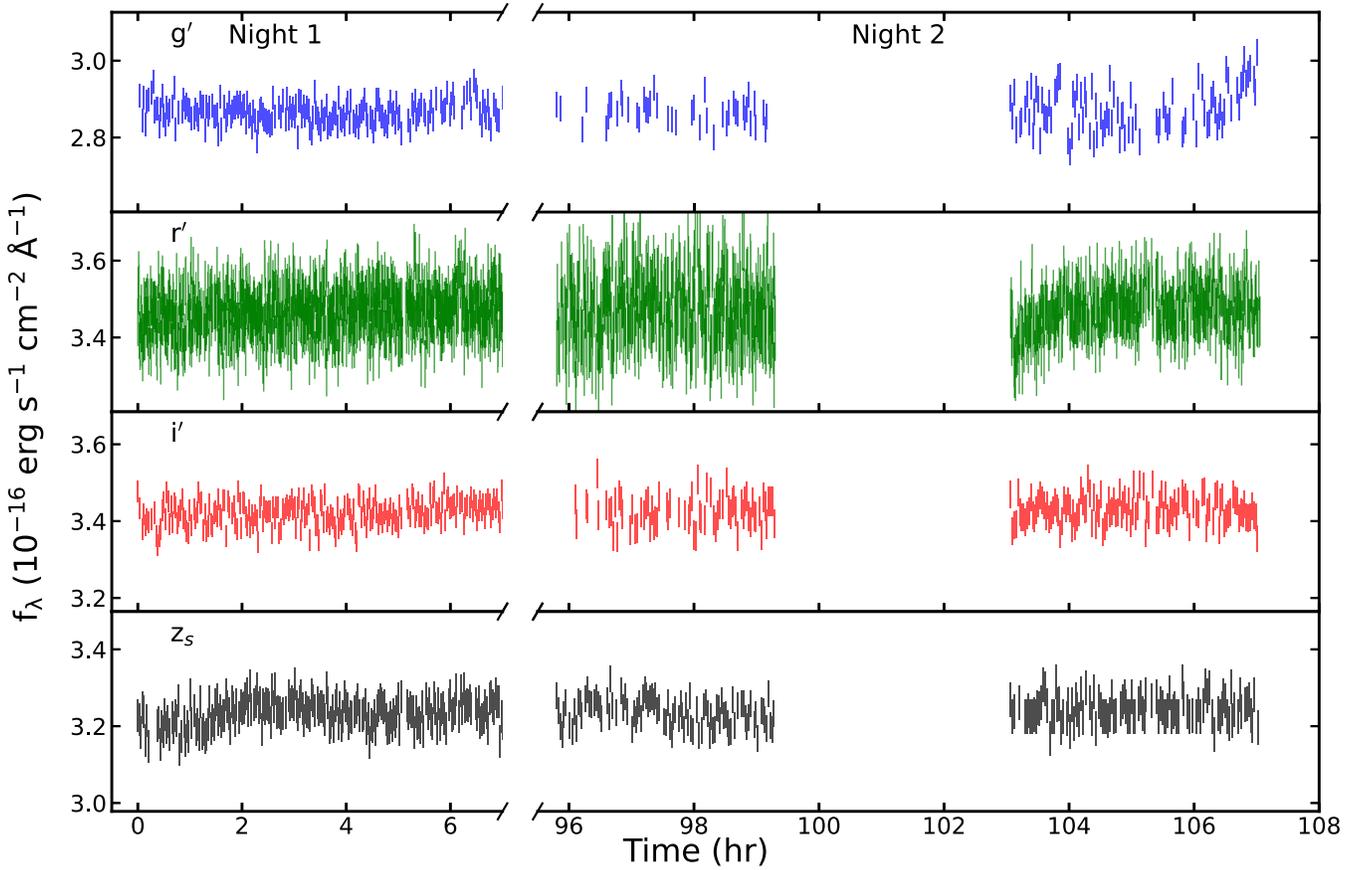

**Figure 4.** Light curves of J0249−0815. Two-night photometry is performed using the MuSCAT four-channel imager. Night 1 and the first half of Night 2 are observed by MuSCAT 3 at the Haleakalā Observatory, Hawaii, while the rest is observed by MuSCAT 4 at Siding Spring Observatory, Australia. The single exposure times of $g'$, $r'$, $i'$, and $z_s$ are 100 s, 25 s, 100 s, and 100 s. Their photometric errors are 1.0%, 1.8%, 0.8%, and 1.3%, respectively. No significant variations were detected during this monitoring period.

**Table 2**
Variability Properties in Multiband Light Curves for Two-night Monitoring with LCOGT

| Filter | $F_{var,N1}$ | $F_{var,N2,1}$ | $F_{var,N2,2}$ |
| --- | --- | --- | --- |
| (1) | (2) | (3) | (4) |
| $g'$ | 0 | 0 | 0.014 ± 0.001 |
| $r'$ | 0 | 0 | 0 |
| $i'$ | 0.003 ± 0.001 | 0 | 0 |
| $z_s$ | 0 | 0 | 0 |

**Note.** Columns (2)–(4) refer to the fractional variability of Night 1 and two different segments of Night 2. $F_{var} = 0$ indicates negative values of $\Delta^2 - \langle\sigma^2\rangle$ in Equation (3).

uncertainties of J0249−0815 are approximately three times larger than that of NGC 4395, with the median photometric error in the g band as around 0.3% and 1.0% for NGC 4395 and J0249−0815, respectively.

Besides, the brightness contrast between the nucleus and the host is the most important factor for variability detection. To estimate the host-AGN contrast for our target, we performed a two-dimensional image decomposition using GALFIT (C. Y. Peng et al. 2002, 2010) based on the Dark Energy Camera Legacy Survey (DECaLS) g-band ($\lambda_{eff}$ = 4730 Å, 1 pixel = 0″.27) image. We use stars within the FOV to model the PSF and employ a constant sky background in the model. During our initial fitting, we used one Sérsic component for a bulge component, an exponential component for an outer disk component, and a PSF for any unresolved component (e.g., a potential AGN component or a star cluster). Adding another Sérsic component for a bar component significantly improved the fit. The best-fit results and parameters are shown in Figure 6 and Table 3. The galaxy light is dominated by the exponential disk, a bulge ($n$ = 0.89) and a bar ($n$ = 0.26), consisting of 96.4% of the total flux. Therefore, both image and spectroscopic analyses (see Figure 2) indicate that the AGN contribution to our source is relatively weak, at approximately 5%, which aligns with the findings from a systematic image decomposition using high-resolution images (Y.-F. Jiang et al. 2011).

Regarding NGC 4395, H. A. G. Cruz et al. (2023) estimated the AGN's contribution to be approximately 30% of the total flux at 5100 Å, by modeling the spectral energy distribution with inputs from prior high-spatial-resolution image decompositions (D. J. Carson et al. 2015). Given that the intranight variability of NGC 4395 is around 10%, this suggests that variability can be detected when the photometric accuracy is better than 3%, a threshold easily met by LCOGT (0.3% for NGC 4395). However, given the AGN's contribution could be less than 5%, detecting variability in our target requires photometric accuracy better than 0.5%. However, the practical accuracy of LCOGT is only ∼1%. This means a telescope with a larger aperture size, e.g., 5–10 m, with sufficient single-epoch exposure, may detect the weak variability of our target.

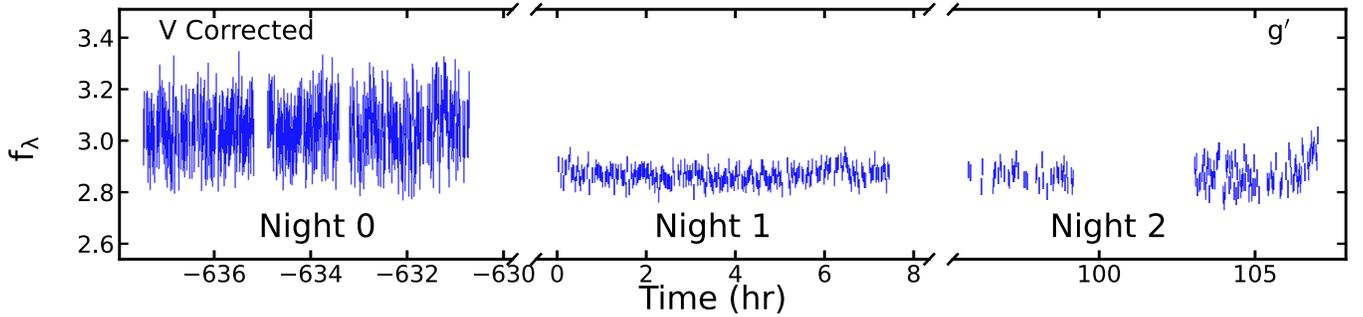

**Figure 5.** Light curves of J0249−0815 on three-night monitoring. The $V$-band light curve is observed by the 1.3 m DFOT in India and the $g'$-band light curve is from LCOGT. The single exposure time of the $V$ band is 50 s. The photometric error in the $V$-band light curve is 3.6%. The filter difference (∼0.406 mag) between $V$ and $g'$ is corrected based on the P200 spectrum. $f_\lambda$ is in units of $10^{-16}$ erg s$^{-1}$ cm$^{-2}$ Å$^{-1}$.

2. *Intrinsically weak variability:* Variability is significant and ubiquitous among luminous quasars (M.-H. Ulrich et al. 1997), where the contribution of host light is negligible due to the large brightness contract. According to previous observations of quasars (e.g., W. Zuo et al. 2012), optical variability generally increases with decreasing AGN luminosity. Moreover, quasar variability is anticorrelated with BH mass given a controlled Eddington ratio, which is consistent with the prediction of thermal fluctuation of the accretion disk model (e.g., M. Sun et al. 2020). These predict that the intrinsic variability amplitude of our target could be slightly smaller on average than that in NGC 4395, considering a comparable Eddington ratio and assuming these correlations still hold in the low-mass and low-luminosity regimes. Therefore, much weak intrinsic variability in our target is not anticipated, although we cannot rule out that it may enter a relatively quiescent state given the weakened broad H$\alpha$ line.

3. *Variability timescale*: Intranight variability is ubiquitous in blazars but rare in normal radio-quiet AGNs (V. Negi et al. 2023). Scaling down the variability timescale to the low-mass regime, IMBHs around $10^5 M_\odot$ are expected to exhibit very fast variation in a few hours. Indeed, NGC 4395, the only RM-confirmed IMBH, has shown significant variations on timescale of hours to years based on the dedicated monitoring (B. M. Peterson et al. 2005; J.-H. Woo et al. 2019; J. W. Montano et al. 2022; I. M. McHardy et al. 2023) and wide-sky surveys (e.g., TESS, Zwicky Transient Facility) (E. C. Bellm et al. 2019; C. J. Burke et al. 2020). Previous observations suggest that, on average, approximately 2–3 hr in NGC 4395 are needed to detect a variable feature (an up-and-down pattern) that is useful for lag detection. For our target, this duration can extend to 20–30 hr after adjusting for a scaling factor of 10 based on BH mass or continuum luminosity, roughly comparable with our two-night LCOGT monitoring period. On the other hand, the predicted damping timescale (a saturation timescale for variability in damped random walk model and possibly linked to the thermal timescale) of an IMBH of $10^{5.5} M_\odot$ is only around 10 days (C. J. Burke et al. 2021), less than our monitoring period of a month. Given both timescales, we would expect to observe some indications of variability trends, if not an entire variable feature. However, it is possible that the intranight variability may be absent during our high-cadence monitoring across two nights, as AGN variability is stochastic and unpredictable.

4. *Intrinsic extinction:* Dust extinction may smooth the variability of the central AGN. To evaluate the extinction from the local host in AGN, we utilized the hydrogen Balmer decrement, which is insensitive to gas temperature and density in low-density, dilute radiation field conditions (D. E. Osterbrock & G. J. Ferland 2006; X. Dong et al. 2008). Specifically, we calculated the line flux ratio between the H$\alpha$ and H$\beta$ lines from the P200 spectrum. Since narrow H$\beta$ is extremely weak and the spectral decomposition results can be significantly biased, we calculated the flux ratios of their narrow components, broad components, and total profiles as 4.58, 2.08, and 2.36 respectively, with a mean of 3.0. This is consistent with the typical Case B recombination value of 2.87 for H II regions photoionized by a hot star, suggesting that dust extinction is not significantly affecting our target.

To prevent similar challenges in future photometric IMBH-RM campaigns, a tailored simulation of IMBH variation in the host galaxy, multiband (e.g., X-ray/radio) confirmations of IMBH nature, and the implementation of a preliminary dual-band test[28] are the keys to purify the IMBH sample and thus for effective variability detection and lag measurement. In addition, verifying the current status of the broad-line component through spectroscopic observations could also help avoid stellar processes induced broad emission lines and turn-off AGN scenario as a changing-look AGN.

## 5. Conclusions

We performed an optical continuum RM campaign targeting the dwarf galaxy J0249−0815 at $z = 0.0297$, serving as a trial before our IMBH-RM campaign, which aims to measure continuum lags, thereby constraining the size of the accretion disk. It includes two nights of multiband, high-cadence monitoring, along with a preliminary one-night single-band test, spanning 1 month from 2023 October 13 to November 13 UT. Despite expectations, we detected no significant continuum variability (less than 1.4%) in our IMBH candidate on timescales of either a few hours or a month. We attribute this to

---

[28] For instance, employing a blue band, such as the $u/g$ band, can enhance the variability detection, as AGN fluctuations are more pronounced at shorter wavelengths. Simultaneously, employing a red band (e.g., $z$ band) would enable the assessment of any potential lags relative to the blue band and facilitate the evaluation of variability attenuation across increasing wavelengths.

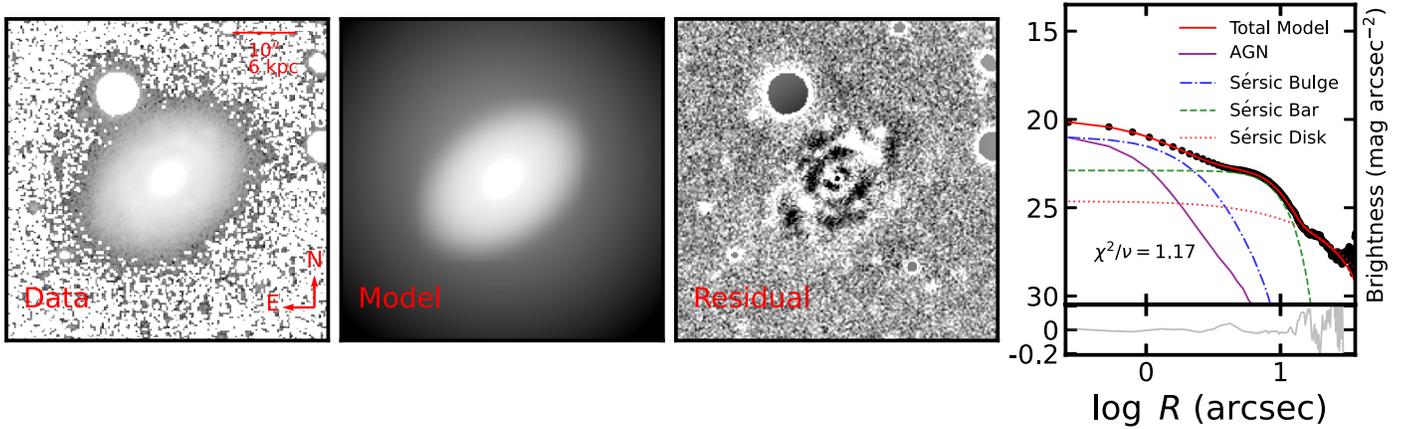

**Figure 6.** Host image decomposition based on the DECaLS $g$-band image using GALFIT. The plots from left to right are the original data, the best-fit model, the fitting residuals (data−model), and the radial surface brightness profile. Three Sérsic profiles account for the central bulge, bar, and outer disk components, respectively. We added a PSF profile to model the potential AGN component. Three stars in the image are masked. Our best-fit parameters are listed in Table 3.

**Table 3**
DECaLS $g$-band Photometric Decomposition Results from GALFIT Analysis

|     | Component       | $m$ (mag)        | $R_e$ (kpc) / $R_s$ (kpc) | $n$              |
|-----|-----------------|------------------|---------------------------|------------------|
| (1) | PSF             | $19.86 \pm 0.01$ |                           |                  |
| (2) | Sérsic bulge    | $18.42 \pm 0.01$ | $0.94 \pm 0.005$          | $0.89 \pm 0.02$  |
| (3) | Sérsic bar      | $16.96 \pm 0.01$ | $4.50 \pm 0.01$           | $0.26 \pm 0.00$  |
| (4) | Exponential disk| $17.52 \pm 0.05$ | $5.96 \pm 0.28$           |                  |

**Note.** Row (1): $g$-band magnitude of the PSF component in our best-fit model; rows (2–3) magnitude, effective radius $R_e$ and Sérsic index of the central bulge and bar components; row (4) magnitude and scale length $R_s$ of the exponential extended disk. The reduced $\chi_\nu^2$ of our best-fit model is 1.17.

the significant contribution of host light, which effectively dilutes the weak variability originating from the central IMBH. This outcome emphasizes several key points for future RM campaigns: selecting targets with a higher AGN-to-host flux ratio to reduce dilution effects; using larger telescopes or more sensitive instruments capable of detecting subtle variability signals from IMBHs; performing dual-band preliminary monitoring and tailored simulations to evaluate IMBH variability in different scenarios; and conducting/collecting complementary observations in other wavelengths (e.g., X-ray, radio) to confirm IMBH activity.

## Acknowledgments

We thank J.Z. Zhu, N. Jiang, and Z.Y. Li for the helpful discussion on the difference image pipeline. We thank M.Y. Zhuang for the helpful discussion on GALFIT parameter setting. We thank E. Manne-Nicholas for help with LOCGT observations. H.X.G. is supported by the National Key R&D Program of China No. 2022YFF0503402, 2023YFA1607903, and Future Network Partner Program, CAS, No. 018GJHZ2022029FN, Overseas Center Platform Projects, CAS, No. 178GJHZ2023184MI. M.F. G. is supported by the Shanghai Pilot Program for Basic Research-Chinese Academy of Science, Shanghai Branch (JCYJ-SHFY-2021-013), the National SKA Program of China (grant No. 2022SKA0120102), the science research grants from the China Manned Space Project with No. CMSCSST-2021-A06, and the Original Innovation Program of the Chinese Academy of Sciences 715 (E085021002). L.C.H. acknowledges support from the NSFC (11991052, 12233001), the National Key R&D Program of China (2022YFF0503401) and the China Manned Space Project (CMS-CSST-2021-A04, CMS-CSST-2021-A06). M.M. acknowledges support from the Spanish Ministry of Science and Innovation through the project PID2021-124243NB-C22. M.M. is also partially supported by the program Unidad de Excelencia María de Maeztu CEX2020-001058-M. I.D., A.B.K., L.C.P., and O.V. acknowledge funding provided by the University of Belgrade, Faculty of Mathematics (contract No. 451-03-66/2024-03/200104), and the Astronomical Observatory Belgrade (contract No. 451-03-66/2024-03/200002) through grants by the Ministry of Science, Technological Development and Innovation of the Republic of Serbia. H.C.F. and S.S.L. are supported by National Natural Science Foundation of China (grant Nos. 12303022 and 12203096). V.U. acknowledges funding support from STScI grant Nos. JWST-GO-01717.001-A, HST-AR-17065.005-A, HST-GO-17285.001-A, and NASA ADAP 80NSSC23K0750.

This paper is based on observations made with the MuSCAT instruments, developed by the Astrobiology Center (ABC) in Japan, the University of Tokyo, and LCOGT. MuSCAT 3 was developed with financial support by JSPS KAKENHI (JP18H05439) and JST PRESTO (JPMJPR1775), and is located at the Faulkes Telescope North in Maui, HI (USA), operated by LCOGT. MuSCAT 4 was developed with financial support provided by the Heising-Simons Foundation (grant 2022-3611), JST grant No. JPMJCR1761, and the ABC in Japan, and is located at the Faulkes Telescope South at Siding Spring Observatory (Australia), operated by LCOGT.

*Facilities:* FTN, FTS, ARIES: DFOT, Hale.

*Software*: Astropy (Astropy Collaboration et al. 2018), Linmix (B. C. Kelly 2007), GALFIT (C. Y. Peng et al. 2002), PyQSOFit (H. Guo et al. 2018).

## Appendix

### A.1. Narrow Emission-line Variation in J0249−0815 Over 20 yr

To evaluate the intrinsic variation of narrow emission lines in a low-luminosity and low-mass AGN, we performed a careful examination. We reexamined the 2D image in spectral extraction and confirmed that the cosmic-ray removal process does not affect the narrow-line region. Our spectral decomposition in Figure 2 shows that the total narrow-line fluxes of [N II]$\lambda$6585, [S II]$\lambda$6732, and [O III]$\lambda$5007 from the P200 spectrum are 11%–20% larger than those from the SDSS spectrum, while the narrow components in the P200 spectrum of H$\alpha$ and H$\beta$ are 30%, 17% smaller than those from the SDSS spectrum. According to previous photoionization calculations (M. Eracleous et al. 1995), [O III]$\lambda$5007 will decay rapidly ($\sim$4 yr) with the decrease of the ionizing flux, while this decay will take $\sim$30 yr for both [N II]$\lambda$6585 and [S II]$\lambda$6732, without accounting for the time delay between the ionizing source and narrow-line-emitting region. This means the [N II] and [S II] line fluxes are not expected to vary with [O III] within a timescale of 20 yr, indicating the narrow-line variation is not intrinsic. We suggest that this phenomenon can be explained by different observational conditions. The P200 observation uses a small slit width of 1.″5, which includes less contamination from the H II region in the host compared to the SDSS fiber size of 3.″0, resulting in lower Balmer line flux yet higher metal lines as expected. Thus, the variations in the narrow lines between the SDSS and P200 spectra are most likely artifacts rather than intrinsic variability.

## ORCID iDs

Wenwen Zuo https://orcid.org/0000-0002-4521-6281
Hengxiao Guo https://orcid.org/0000-0001-8416-7059
Jingbo Sun https://orcid.org/0000-0003-3472-4392
Qi Yuan https://orcid.org/0000-0003-4671-1740
Paulina Lira https://orcid.org/0000-0003-1523-9164
Minfeng Gu https://orcid.org/0000-0002-4455-6946
Philip G. Edwards https://orcid.org/0000-0002-8186-4753
Alok C. Gupta https://orcid.org/0000-0002-9331-4388
Shubham Kishore https://orcid.org/0000-0001-8716-9412
Jamie Stevens https://orcid.org/0000-0002-5841-3348
Tao An https://orcid.org/0000-0003-4341-0029
Zhen-Yi Cai https://orcid.org/0000-0002-4223-2198
Hai-Cheng Feng https://orcid.org/0000-0002-1530-2680
Luis C. Ho https://orcid.org/0000-0001-6947-5846
Dragana Ilić https://orcid.org/0000-0002-1134-4015
Andjelka B. Kovačević https://orcid.org/0000-0001-5139-1978
Sha-Sha Li https://orcid.org/0000-0003-3823-3419
Mar Mezcua https://orcid.org/0000-0003-4440-259X
Luka Č. Popović https://orcid.org/0000-0003-2398-7664
Mouyuan Sun https://orcid.org/0000-0002-0771-2153
Tushar Tripathi https://orcid.org/0009-0006-3586-2489
Vivian U https://orcid.org/0000-0002-1912-0024
Oliver Vince https://orcid.org/0009-0008-5761-3701
Jianguo Wang https://orcid.org/0000-0003-4156-3793
Junxian Wang https://orcid.org/0000-0002-4419-6434
Shu Wang https://orcid.org/0000-0002-2052-6400
Xuebing Wu https://orcid.org/0000-0002-7350-6913
Zhenya Zheng https://orcid.org/0000-0002-9634-2923